\documentclass[aps,prl,twocolumn,showpacs,superscriptaddress]{revtex4-2}

\usepackage{amsmath}
\usepackage{graphicx}
\usepackage{lmodern}
\usepackage{amsmath}

\usepackage{color}
\usepackage{amssymb}
\usepackage{bm}
\usepackage{braket}
\usepackage{mathtools}

\usepackage[dvipsnames]{xcolor}
\usepackage[colorlinks=true]{hyperref}
\hypersetup{
    colorlinks=true,
    linkcolor=blue,
    citecolor=blue,
    urlcolor=blue
} 

\makeatletter
\def\maketitle{
\@author@finish
\title@column\titleblock@produce
\suppressfloats[t]}
\makeatother
\begin{document}

\title{Exact deconfined gauge structures in the higher-spin Yao-Lee model: \\a quantum spin-orbital liquid with spin fractionalization and non-Abelian anyons}

\author{Zhengzhi Wu}
\thanks{These two authors contributed equally to this work.}
\affiliation{Institute for Advanced Study, Tsinghua University, Beijing 100084, China}

\author{Jing-Yun Zhang}
\thanks{These two authors contributed equally to this work.}
\affiliation{Institute for Advanced Study, Tsinghua University, Beijing 100084, China}

\author{Hong Yao}
\email{yaohong@tsinghua.edu.cn}
\affiliation{Institute for Advanced Study, Tsinghua University, Beijing 100084, China}

\begin{abstract}
 The non-integrable higher spin Kitaev honeycomb model has an exact $\mathbb{Z}_2$ gauge structure, which exclusively identifies quantum spin liquid (QSL)
 in the half-integer spin Kitaev model. But its constraints for the integer-spin Kitaev model are much limited, and even trivially gapped insulators cannot be excluded. The physical implications of exact $\mathbb{Z}_2$  gauge structure, especially $\mathbb{Z}_2$ fluxes, in integer-spin models remain largely unexplored. In this Letter, we theoretically show that a spin-S Yao-Lee model (a spin-orbital model with SU(2) spin-rotation symmetry) possesses a topologically-nontrivial quantum spin-orbital liquid (QSOL) ground state for \textit{any} spin (both integer and half-integer spin) by constructing exact deconfined fermionic $\mathbb{Z}_2$ gauge charges. We further show that the conserved $\mathbb{Z}_2$ flux can also demonstrate the intriguing spin fractionalization phenomena in the nonabelian topological order phase of the spin-1 Yao-Lee model. Its deconfined $\mathbb{Z}_2$ vortex excitation carries fractionalized spin-$\frac{1}{2}$ quantum number in the low-energy subspace, which is also an nonabelian anyon. 
 Our exact manifestation of spin fractionalization in an integer-spin model is rather rare in previous studies, and is absent in the Kitaev honeycomb model.  
\end{abstract}
\date{\today}

\maketitle

{\it Introduction.}---Quantum spin liquids (QSL) \cite{ANDERSON1973153, doi:10.1126/science.1163196,Balents2010,Savary_2017,RevModPhys.89.025003,doi:10.1146/annurev-conmatphys-031218-013401,doi:10.1126/science.aay0668,2004qftm,Fradkin_2013,Sachdev_2023} are long-range entangled quantum phases beyond the Landau paradigm \cite{PhysRevLett.96.110405,PhysRevLett.96.110404,RevModPhys.89.041004,anderson1987resonating,KRS1987,PhysRevLett.61.2376,PhysRevLett.86.1881} and also promising candidates for quantum computation \cite{KITAEV20032,Nayak-RMP2008}. They have attracted a great amount of studies on frustrated spin models \cite{Read-Sachdev1991,XGWen1991,PhysRevB.65.165113,PhysRevB.72.104404,doi:10.1126/science.1201080,PhysRevLett.118.137202,PhysRevB.86.024424,PhysRevLett.121.107202,Sorella2012,PhysRevX.11.031034,PhysRevX.12.041029,PhysRevB.91.075112,PhysRevB.100.165123,PhysRevX.7.031020,PhysRevX.10.021042,PhysRevLett.121.057202,LIU20221034,PhysRevX.12.031039,song2019unifying,PhysRevX.10.011033,qian2023absence}. It is remarkable that the spin-$\frac{1}{2}$ Kitaev honeycomb model \cite{Kitaev2006} is exactly solvable with QSL ground states, and there have been increasing experimental evidences that Kitaev QSL can be possibly realized in materials \cite{PhysRevLett.102.017205,doi:10.1146/annurev-conmatphys-031115-011319,doi:10.1146/annurev-conmatphys-033117-053934,TREBST20221}. Besides spin degrees of freedom (DOF), localized electrons in many Mott insulators also have orbital DOF. Topological liquid states with both spin and orbital DOF are often called QSOL. Moir\'{e} systems are possible experimental platforms to realize QSOL \cite{PhysRevLett.125.117202,PhysRevLett.127.247701,PhysRevLett.121.087001} and  numerical evidence of QSOL has been reported in studies on various SU(4) Heisenberg models and Kugel-Khomskii models \cite{PhysRevLett.121.087001,PhysRevX.2.041013, PhysRevB.100.205131,PhysRevB.107.L180401,PhysRevB.80.064413,PhysRevLett.125.117202,PhysRevLett.127.247701,zhang2023variational,JIN2022918}. Moreover, generalizations of the Kitaev spin-1/2 model have lead to various exactly solvable models with QSOL ground states  \cite{Yao2011,Yao2009,Nussinov2009,Wu2009,Chua2011,PhysRevB.102.075110,PhysRevB.79.214440,PhysRevB.108.075111}. 
In particular, the spin-$\frac{1}{2}$ Yao-Lee model \cite{Yao2011} is an exactly solvable spin-orbital model with SU(2) spin-rotation symmetry. Recently, there have been increasing research interests on the spin-$\frac{1}{2}$ Yao-Lee model and its various extensions  \cite{PhysRevB.103.075144, PhysRevLett.129.177601, PhysRevLett.127.127201,PhysRevLett.125.257202,PhysRevLett.125.067201,PhysRevB.106.125144,PhysRevB.102.201111,PhysRevB.103.075144,PhysRevB.108.224427,PhysRevB.104.L060403,PhysRevB.98.155105,PhysRevB.103.155160,Poliakov2023,Polyakov2023,PhysRevB.84.235148,seifert2024wegners,mandal2024nonhermitian}, which have revealed unexpected fertile physics. Moreover, a microscopic roadmap to a spin-1/2 Yao-Lee spin-orbital liquid was proposed in Ref. \cite{Kee2024}. 

More recently, higher-spin Kitaev models have attracted numerous research interests \cite{PhysRevB.98.214404,PhysRevB.78.115116,doi:10.7566/JPSJ.87.063703,PhysRevResearch.2.022047,doi:10.7566/JPSCP.30.011086,PhysRevResearch.2.033318,PhysRevB.104.024417,PhysRevResearch.3.013160,PhysRevB.105.L060403,PhysRevB.105.L060405,PhysRevB.99.104408,PhysRevResearch.4.013205,Jin2022,PhysRevB.108.075111,Rousochatzakis2018,PhysRevLett.130.156701,liu2023symmetries}, partly motivated by their possible experimental realizations \cite{Xu2018,PhysRevLett.123.037203,PhysRevLett.124.087205,PhysRevResearch.3.013216,PhysRevB.104.184415,PhysRevB.107.014411}. 
An exact $\mathbb{Z}_2$ gauge structure exists for any spin-S \cite{PhysRevLett.130.156701,liu2023symmetries}, which powerfully identifies the QSL in the half-integer spin Kitaev model without solving it. But it imposes almost no constraints on the ground states of integer-spin Kitaev models. 
Indeed, the ground state is a trivial product state in the anisotropic coupling regime, and so far no QSL has been obtained in the integer spin Kitaev models in analytically controlled ways \cite{PhysRevB.99.104408,PhysRevLett.130.156701}. Numerical results about their nature are still controversial \cite{PhysRevB.102.121102,PhysRevResearch.2.033318}. As a result, it is an outstanding open question 
whether QSL can be reliably identified in integer spin models by solely referring to the exact $\mathbb{Z}_2$ gauge structure. Moreover, previous studies on the diagnosis of QSL mainly investigate the $\mathbb{Z}_2$ gauge charges \cite{PhysRevLett.130.156701,liu2023symmetries}, while the physical implications of conserved $\mathbb{Z}_2$ fluxes, and their connections with other important physical consequences of QSL, such as spin fractionalization, remains \textit{terra incognita.} Deconfined spinon excitation with fractionalized spin quantum number has not yet been reported in the spin-S Kitaev model, and it is especially challenging to exactly demonstrate its existence in 2+1d integer spin models due to the weaker spin quantum fluctuation than half-integer spin models.

In this work, we address the above questions by showing that the spin-S Yao-Lee model always has nontrivial QSOL for any value of spin-S, since it always has exact deconfined $\mathbb{Z}_2$ fermionic gauge charges. This is 
shown both through an exact Majorana parton construction 
inspired by the spin-S Kitaev model \cite{PhysRevLett.130.156701} and through Levin-Wen 'T-shape' fermion exchange 
method \cite{PhysRevB.67.245316}. 
Moreover, we use analytically controlled methods to show that the spin-1 Yao-Lee model can host gapless QSOL or even nonabelian topological order. Its deconfined $\mathbb{Z}_2$ vortex excitation is a nonabelian anyon with fractionalized spin-$\frac{1}{2}$ quantum number in the low energy subspace, which demonstrates spin fractionalization in the ground state. This spinon excitation is exact since the  $\mathbb{Z}_2$ flux is conserved. Remarkably, the nonabelian topological order here is the fascinating spin-1 bosonic Moore-Read Pfaffian topological order \cite{GREITER1992567,PhysRevLett.91.030402,PhysRevLett.87.120405,PhysRevLett.107.146803,PhysRevB.89.165125,PhysRevLett.102.207203,Glasser_2015,PhysRevB.106.115131,PhysRevB.97.195158,PhysRevB.103.075130,PhysRevB.98.184409,PhysRevB.95.140406,PhysRevB.91.245126}, and our model provides a 2+1d local spin Hamiltonian sought after in previous numerical studies \cite{PhysRevLett.102.207203,Glasser_2015,PhysRevB.106.115131,PhysRevB.97.195158,PhysRevB.103.075130,PhysRevB.98.184409,PhysRevB.105.155104}. 
We further investigate the SU(2) symmetric Yao-Lee model in the large-S limit where quite rich spin physics are revealed.

\begin{figure}
  \includegraphics[width=6cm]{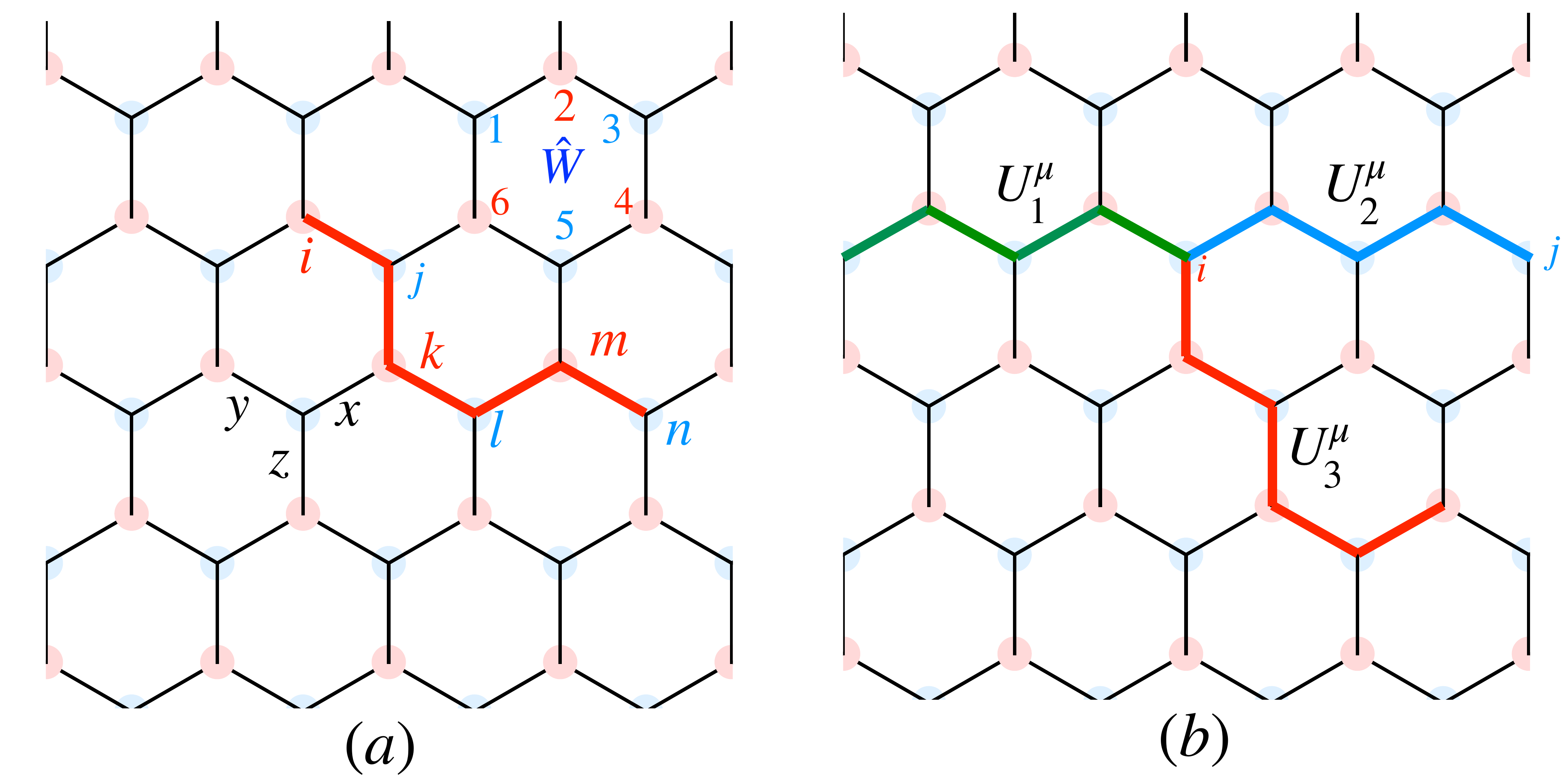}
 \caption{(a) Operator $\hat{W}$ is the flux operator. The red line represents the open string operator. (b) The support of three movement operators $U^{\mu}_1,U^{\mu}_2,U^{\mu}_3$ of the spin-S spin-orbital model \eqref{ham:yl} on the lattice. At the ends of each open string operator $U^{\mu}$, there are two excitations which have fermionic statistics and can be moved by $U^{\mu}$. }
 \label{fig:lat}
\end{figure}

{\it The model.}---In this Letter, we mainly focus on the following spin-S Yao-Lee model on the honeycomb lattice: 
\begin{equation}
\hat{H}=-\sum_{\langle i j\rangle \in \mu} J_{\mu} [\vec{S}_i\cdot \vec{S}_j]\otimes[ \tau_{i}^{\mu} \tau_{j}^{\mu}],
\label{ham:yl}
\end{equation}
where $\vec{S}_i$ are spin-S operators on site $i$, and the Pauli matrices $\tau_{i}^{\mu}$ represent the orbital DOF (the case of two orbitals). 
The spin-spin interactions are nearest-neighbour Heisenberg interactions with SU(2) spin-rotational symmetry, and the orbital interactions are similar to those in the Kitaev honeycomb model, which features Ising couplings $\tau_{i}^{\mu} \tau_{j}^{\mu}$ with coupling constants $J_{\mu}$ in the $\mu=x,y,z$ types of bonds $\langle ij\rangle$, as shown in Fig. \ref{fig:lat}.

This model is exactly solvable when $S=\frac{1}{2}$ \cite{Yao2011}. 
Here we give a brief review of this minimal model. The Pauli operators $\vec{\sigma}_i=2\vec{S}_i$ and $\vec \tau_i$ can be written in terms of Majorana fermions: $\vec{\sigma}_{i}=-\frac{i}{2}\vec{d}_i\times \vec{d}_i$ and $\vec{\tau}_{i}=-\frac{i}{2}\vec{c}_i\times \vec{c}_i$ with 
projection: 
$id^x_id^y_id^z_ic^x_ic^y_ic^z_i=1$. The spin-$\frac{1}{2}$ Yao-Lee model becomes 
$\hat{H}_{S=\frac{1}{2}}^f=-\frac{J_{\mu}}{4}\sum_{\langle i j\rangle \in \mu}  u_{\langle ij\rangle} i\vec{d}_i\cdot\vec{d}_j$, where $u_{\langle ij\rangle}=-ic_i^{\mu}c_j^{\mu}$ are static $\mathbb{Z}_2$ gauge fields. Its ground state is a QSOL without symmetry breaking for any coupling constants $J_{\mu}$, which has $Z_2$ topological order in the anisotropic regime, and becomes gapless QSOL in the isotropic regime. 
In the following, we show that the ground state of the Yao-Lee model  
is always a nontrivial QSOL for any value of spin-S by constructing deconfined fermionic $\mathbb{Z}_2$ gauge charges.

{\it Deconfined fermions in the higher spin Yao-Lee model.}---We now construct the exact $\mathbb{Z}_2$ gauge structures of the spin-S Yao-Lee model through a parton construction.
First, we decompose the spin-$S$ operator $\vec{S}_i$ into 2$S$ spin-$\frac{1}{2}$ Pauli matrices similar to the trick used in the spin-$S$ Kitaev honeycomb model \cite{PhysRevLett.130.156701}: $\vec{S}_i=\frac{1}{2}\sum_{a=1}^{2S}\vec{\sigma}_{a,i}$, together with the projection to leave only the spin-S physical Hilbert space: $(\frac{1}{2}\sum_{a=1}^{2S}\vec\sigma_{a,i})^2=S(S+1)$. Then we take the following Majorana partons representation of all the Pauli matrices  $\vec{\tau}_i$ and $\vec{\sigma}_{a,i}$:
\begin{equation}
\begin{aligned}
    &\vec{\tau}_{i}=-\frac{i}{2}\vec{c}_i\times\vec{c}_i,\quad \vec{\sigma}_{1,i}=-\frac{i}{2}\vec{d}_{1,i}\times\vec{d}_{1,i},\\
    &\vec{\sigma}_{a,i}= i  \vec{d}_{a,i}d_{a,i}^0,\quad  a=2,3,\cdots,2S.
    \end{aligned}
\end{equation}
This parton construction returns to the Majorana representation of the spin-$\frac{1}{2}$ Yao-Lee model \cite{Yao2011} when $S=\frac{1}{2}$. We adopt this construction since 
$\mathbb{Z}_2$ gauge charges only consist of $\vec d_{i}$ Majorana fermions coming from the fractionalization of spin operators $\vec{S}_i$, as we will immediately show below. This implies the spin DOF 
can also fractionalize in the QSOL, which is qualitatively different from the pure orbital Kitaev model and has richer physics.

Using this Majorana representation above, the spin-S Yao-Lee model $\hat{{H}}$ can be written as: 
\begin{equation}
\begin{aligned}
   \hat{H}^f=-\frac{1}{4}\sum_{\langle i j\rangle \in \mu}  J_{\mu}\hat{u}_{\langle ij\rangle}\sum_{a,b=2}^{2S}i{\vec D}_{a,i}\cdot{\vec D}_{b,j},  
   \end{aligned}
\end{equation}
where $\vec{D}_{a,i}=\vec{d}_{1,i}+\vec{d}_{a,i}d_{1,i}^{x}d_{1,i}^{y} d_{1,i}^{z}d_{a,i}^0$ is the composite field with $a=2,\cdots,2S$. The $\mathbb{Z}_2$ gauge fields $\hat{u}_{\langle ij\rangle}=-ic^{\mu}_{i}c_j^{\mu}$ are conserved quantities: $
\left[\hat{u}_{\langle i j \rangle}, \hat{H}^f\right]=0$ and $\left[\hat{u}_{\langle i j \rangle}, \hat{u}_{\langle i^{\prime} j^{\prime}\rangle}\right]=0
$. The conserved flux operator can be  written in physical operators: $
W_{p}=\tau_{1}^{y} \tau_{2}^{z} \tau_{3}^{x} \tau_4^{y} \tau_5^{z} \tau_6^{x}
$, as is illustrated in Fig. \ref{fig:lat}(a).

Now we consider an open string operator, e.g. on the string shown in Fig. \ref{fig:lat}(a): 
\begin{equation}
    \begin{aligned}
        U^{\mu}&=e^{i\pi S_{i}^{\mu}}\tau_{i}^{y} \tau_{j}^{x}\tau_{k}^{x} \tau_{l}^{z} \tau_{m}^{z}\tau_{n}^{y}e^{i\pi S_{n}^{\mu}}\\
        &\propto \Gamma_i^{\mu}(\hat{u}^y_{\langle ij\rangle}\hat{u}^z_{\langle jk\rangle}\hat{u}^y_{\langle kl\rangle}\hat{u}^x_{\langle lm\rangle}\hat{u}^y_{\langle mn\rangle})\Gamma_n^{\mu},
    \end{aligned}
\end{equation}
where the '$\propto$' in the second line of the equation above only means there can be a path and spin-S dependent $U(1)$ phase in the parton representation. 
And $\Gamma_i^{\mu}=d_{1,i}^{\mu}\Pi_{a=2}^{2S}(id_{a,i}^{\mu}d_{a,i}^{0})$ are the fermionic $\mathbb{Z}_2$ gauge charges at the ends of the open string. 
The fermionic gauge charges $\Gamma^{\mu}$ are always deconfined regardless of the value of spin-S, since $U^{\mu}$ commutes with $\hat{H}$ except at the two ends of the string such that the energy cost is $O(1)$ when two $\Gamma^{\mu}$ are separated to infinity. 
As a result, we immediately arrive at the conclusion that the spin-S Yao-Lee model $\hat{H}$ in Eq.\eqref{ham:yl} always has the nontrivial QSOL ground state for all spin-S~!

The fermionic nature of the gauge charges can also be verified in a more fundamental and parton-representation independent way. 
We can check the statistics of the excitations at the ends of the string using only physical operators, following the approach in \cite{PhysRevB.67.245316,PhysRevB.101.115113}.  
As is illustrated in Fig. \ref{fig:lat}(b), we suppose there are two excitations located at the sites $i$ and $j$ in the initial state, and we then apply two independent movement sequences $U^{\mu}_1U^{\mu}_2U^{\mu}_3$ and $U^{\mu}_3U^{\mu}_2U^{\mu}_1$ to it, respectively. 
The two final states only differ by an exchange of the two excitations, so their wave functions will only differ by a statistical phase $\phi$ of the excitations. 
This is reflected in the algebraic relation of the two movement sequence operators: $U^{\mu}_1U^{\mu}_2U^{\mu}_3=e^{i\phi}U^{\mu}_3U^{\mu}_2U^{\mu}_1$. Here $e^{i\phi}=-1$ for all values of spin-S, which means the excitations at the end points of $U^{\mu}$ obey fermionic statistics.

We continue to analyze its physical properties by finding certain solvable limits. 
In the following, we shall focus on the easy-axis limit of this model for the case of spin-1. Potential material candidates for the spin-1 Yao-Lee model include vanadate compounds with $V^{3+}$ ions where two electrons partially filling the $t_{2g}$ orbitals can realize spin-1 and orbital spin-$\frac{1}{2}$ DOF \cite{PhysRevB.75.184434, PhysRevLett.100.167205}. 
This is the simplest case when the model is not exactly solvable, but it is also very intriguing since the spin-1 Kitaev model does not exhibit spin liquid ground state within analytically controlled solvable limit. 

{\it Easy-axis spin-1 Yao-Lee model.}---We consider the  model in Eq. \eqref{ham:yl} with spin-1 and $J_\mu=J$ in the spin easy-axis anisotropy limit : 
\begin{eqnarray}\label{ham:eal}
    \hat{H}_{S=1}=-J\sum_{\langle ij\rangle \in \mu}[S_i^zS_j^z+a(S_i^xS_j^x+S_i^yS_j^y)]\otimes[ \tau_{i}^{\mu} \tau_{j}^{\mu}],~~~
\end{eqnarray}
where $0\!<\!a\!\ll\! 1$ such that the spin coupling in z-direction is much larger than those in x- and y- directions. 
Here we focus on the case with spatial isotropic couplings $J_\mu=J$, where it is possible to obtain a QSOL different from the familiar Abelian $\mathbb{Z}_2$ topological order, such as gapless QSOL or even non-Abelian topological order. 
Meanwhile, the spatial isotropic coupling limit is the only known regime where spin-1 Kitaev model may have a spin-liquid ground state, but its nature is still under debate \cite{PhysRevB.102.121102,PhysRevResearch.2.033318}. 
It is then 
desired to have an analytically controlled study of the spin-1 model in Eq. \eqref{ham:eal}.

We first derive a low-energy effective Hamiltonian of the model $\hat{H}_{S=1}=\hat{H}_0+a\hat{V}$ through a degenerate perturbation up to the order $a^2$. 
Here the unperturbed part $\hat{H}_0=-J\sum_{\langle i j\rangle \in \mu}S_i^zS_j^z\otimes \tau_{i}^{\mu} \tau_{j}^{\mu}$ is exactly solvable, since all $S_i^z\in\{0,\pm 1\}$ are conserved and we denote the Hamiltonian $\hat{H}_0$ with a fixed $\{S_i^z\}$ configuration as $\hat{H}_0(\{S_i^z\})$.   
After writing the orbital Pauli matrices $\vec{\tau}_i$ into Majorana fermions: $\vec{\tau}_i=i\vec{c}_ic_i^0$,  $\hat{H}_0$ becomes a quadratic fermion Hamiltonian: $\hat{H}_{0}^f=-J\sum_{\langle i j\rangle \in \mu}u_{\langle ij\rangle}ic_i^0c_j^0$, where $u_{\langle ij\rangle}=-ic_i^{\mu}c_i^{\mu}S_i^z S_j^z$ are static $\mathbb{Z}_2$ gauge fields. 
Due to the $\mathbb{Z}_2$ gauge symmetry, an immediate conclusion is that the energy of the ground states of all $\hat{H}_0(\{S_i^z\neq 0\})$ are equal, since these states can be connected to each other through $\mathbb{Z}_2$ gauge transformations in the Majorana representation.  We denote the set of these states as $\mathcal{G}$ and it can be proved that the ground state subspace of $\hat{H}_0$ is $\mathcal{G}$ \cite{SupMat}.  Physically, this means that all the ground states to the zeroth order have one Dirac cone and the inclusion of any site with $S^z_i=0$ will effectively generate a lattice vacancy on that site which increases the energy.

In the ground state sector $\mathcal{G}$ of $\hat{H}_0$, $S_i^z$ can only take two values $\pm 1$, which is like an effective spin-$\frac{1}{2}$ degrees of freedom. 
We denote this effective spin-$\frac{1}{2}$ as $\hat{P}S^z_i\hat{P}=\sigma^z_i$, where $\hat{P}$ is the projection to the low-energy ground state sector $\mathcal{G}$. 
A nontrivial effective Hamiltonian can lift the degeneracy of $\mathcal{G}$ at the second order of $a$. 
As the energy gaps of low-lying intermediate states are approximately the same ($\propto |J|$), we can write down the effective Hamiltonian as \cite{SupMat}:
\begin{equation}
    \hat{H}_{\text{eff}}=\hat{P}\hat{H}_0\hat{P}-J_{\text{eff}}\sum_{\langle ij\rangle}\left(\sigma_i^x\sigma_j^x+\sigma_i^y\sigma_j^y-\sigma_i^z\sigma_j^z\right),
\end{equation}
where $J_{\text{eff}}\propto a^2|J|$. For this effective spin-$\frac{1}{2}$ spin-orbital model above, we can use the same Majorana representation as the spin-$\frac{1}{2}$ Yao-Lee model:
\begin{eqnarray}
    &&\hat{H}_{\text{eff}}^f=-J\sum_{\langle i j\rangle}  \hat{u}_{i j}\left(i d_{i}^{z} d_{j}^{z}\right)\nonumber\\
    &&+2J_{\text{eff}}\sum_{\langle i j\rangle }\left[-i d_{i}^{z} d_{j}^{z}(f^{\dagger}_if_j+H.c.)+2(\hat{n}_i-\frac{1}{2})(\hat{n}_j-\frac{1}{2})\right],~~~
    \label{ferm}
\end{eqnarray}
where $f_i=\frac{1}{2}(d^x_{i}-id^y_i)$ or $-\frac{i}{2}(d^x_{i}-id^y_i)$ if the site $i$ resides on the A or B sublattice, and $\hat{n}_i=f^{\dagger}_if_i$. $\hat{u}_{\langle ij\rangle}=-i c^{\mu}_ic^{\mu}_j$ are static $\mathbb{Z}_2$ gauge fields. Since  the flux gap is proportional to $|J|$, so we can safely take $\hat{u}_{\langle ij\rangle}=1$. 

Although $\hat{H}_{\text{eff}}^f$ in Eq.~\eqref{ferm} is not directly exactly solvable, we can reliably obtain its ground state properties through a mean-field approximation due to the small parameter $a$. The ground state of $\hat{H}_0$ has a Dirac cone with a vanishing density of states, and since $J_{\text{eff}}\ll |J|$, we do not expect this weak four-fermion interaction will induce an symmetry-breaking instability and gap out the Dirac cone. 
So we propose that the ground state wave function of $\hat{H}_{\text{eff}}^f$ in Eq.~\eqref{ferm} can be approximated as: $|\psi_G\rangle=|d^z\rangle\otimes |f\rangle$, where $|d^z\rangle$ and $|f\rangle$ are the $d^z$-fermion and $f$-fermion part, respectively. $|d^z\rangle$ is approximately the ground state of the tight-binding model $\hat H_{0,d}=-J\sum_{\langle ij\rangle} id^z_id^z_j$, and $|f\rangle$ is the ground state of the spinless fermion $t$-$V$ model on the honeycomb lattice with $t=2\lambda J_\text{eff}$ and $V=4J_\text{eff}$. Here $\lambda=\langle i d_{i}^{z} d_{j}^{z}\rangle_d$ is the expectation value of nearest-neighbour Majorana hopping in the state $|d^z\rangle$.
Numerically-exact quantum Monte Carlo studies of the honeycomb spinless $t$-$V$ model have reported a phase transition from Dirac semimetal to charge-density-wave (CDW) order with a critical $V/t$: $(V/t)_c\approx 1.35$  \cite{Wang_2014, PhysRevB.91.241117, Li_2015}. As the effective ratio $\frac{V}{t}$ here is $\frac{V}{t}=\frac{2}{\lambda}>2$,  which means the $f$-fermion is in the CDW phase, which, after translating back to the physical degrees of freedom  equivalently, implies that the spin exhibits a Neel order with opposite spin moments on different sublattices of the honeycomb lattice. 
As a result, the easy-axis spin-1 Yao-Lee model in Eq.~\eqref{ham:eal} has the gapless QSOL ground state with a Dirac cone accompanying with the Neel order in the z-direction.

Physically, the spin order in the easy-axis limit is due to the fact that the spin quantum fluctuation is largely suppressed for small $a$. 
One way to enhance the spin fluctuation is to add a tiny competing ferromagnetic Ising coupling of the same order as $J_{\text{eff}}$ to the model in Eq.~\eqref{ham:eal}: $\hat H'_{S=1}=\hat H_{S=1}+ \hat{H}_{\text{FM}}$, where $\hat{H}_{\text{FM}}=-J_{z}\sum_{\langle ij\rangle}S_i^zS_j^z$ with $J_z>0$. Now the density repulsion in the corresponding $\hat{H}_\text{MF}$ is suppressed to  $V^{\prime}=4J_{\text{eff}}-2J_{z}$, when $\frac{V^{\prime}}{2\lambda J_\text{eff}}$ is smaller than the critical ratio $(\frac{V}{t})_c\approx 1.35 $, the spin order disappears and the ground state will have three Dirac cones. Remarkably, this phase transition is a fractionalized criticality belonging to the Gross-Neveu-$\mathbb{Z}_2^*$ universality class. Specifically, when we take $J_z=2J_{\text{eff}}$ in $\hat{H}'_{S=1}$ , the density repulsion $V^{\prime}$ of the $f$-fermions in the Majorana representation is zero up to $a^2$, so the ground state must lie in the gapless QSOL phase with three Dirac cones. 
Then, the spin-spin correlation $\langle S^z_i S^z_j\rangle \sim |{\bf r}_i-{\bf r}_j|^{-4}$ now has power law decay as $|{\bf r}_i-{\bf r}_j|\rightarrow \infty$, which implies spin also fractionalizes in the QSOL ground state. Note that this phase is not a direct product of spin and orbital DOF since the orbital bond-bond correlation $\langle \tau_i^{\mu}\tau_{j}^{\mu}\tau_{i'}^{\mu}\tau_{j'}^{\mu}\rangle$ contains an extra power-law decaying spin bond-bond correlation $\langle \sigma_i^{z}\sigma_{j}^{z}\sigma_{i'}^{z}\sigma_{j'}^{z}\rangle$ besides the power-law correlations of $d^z$ fermions, which is evident in the parton representation 
\footnote{Under the parton representation, the orbital bond-bond correlation $\langle (\tau^{\mu}_i\tau^{\mu}_{j})(\tau^{\mu}_{i'}\tau^{\mu}_{j'})\rangle$ with nearest-neighbour $\mu$-type bonds $\langle ij\rangle_\mu$ and $\langle i^{\prime}j^{\prime}\rangle_\mu$ becomes: $\langle (\tau^{\mu}_i\tau^{\mu}_j)(\tau^{\mu}_{i^{\prime}}\tau^{\mu}_{j^{\prime}})\rangle\propto \langle \sigma^z_i\sigma^z_j \sigma^z_{i'}\sigma^z_{j'}\rangle\langle (id^z_id^z_j)(id^z_{i^{\prime}}d^z_{j^{\prime}})\rangle$. 
For the spin-orbital liquid phase with three Dirac cones, the spin-spin correlation $\langle \sigma^z_i\sigma^z_j\sigma^z_{i'}\sigma^z_{j'}\rangle$ exhibits a power law decay with distance $|i-i'|$,
the orbital bond-bond correlation is then qualitatively different from that in the phase with one Dirac cone 
since the former is qualitatively affected by the spin-spin power-law correlation. Consequently, the gapless phase with three Dirac cones must be a phase 
with entangled orbital and spin DOF, instead of a direct product of two DOF.}.
In below, we give a more direct manifestation of spin fractionalization by showing that the $\mathbb{Z}_2$ vortex forms a projective representation of $U(1)_z$ spin rotation symmetry if these Dirac cones are gapped.

{\it Non-Abelian topological order.}---Having identified the gapless QSOL ground state of the easy-axis spin-1 Yao-Lee model, we further show that non-abelian topological order can also be realized if time-reversal symmetry (TRS) is spontaneously or explicitly broken in this model.

First, the TRS can be spontaneously broken if we consider the model $\hat{H}'_{S=1}$ on the decorated honeycomb lattice (also known as 
star lattice) \cite{PhysRevLett.99.247203}, with each site in the honeycomb lattice replaced with three sites of a triangle. 
For simplicity, we set all 
bond couplings to be the same $J=1$. 
Its low energy state sector and the effective Hamiltonian are similar to those on the honeycomb lattice, but now the leading order $\hat{H}^0$ gaps $d^z$ fermions with nonzero Chern number $\nu=\pm1$. 
As a result, the weak four-fermion interaction in Eq.~\eqref{ferm} which couples $d^z$ with $f$ fermions must be irrelevant, and $|\psi_G\rangle=|d^z\rangle\otimes |f\rangle$ is still a good approximation of the ground state before projection. 
$|f\rangle$ is the ground state of the 
mean field Hamiltonian $\sum_{\langle i j\rangle }\langle i d_{i}^{z} d_{j}^{z}\rangle_d\left(if^{\dagger}_if_j+H.c.\right)$ with $f_i=\frac{1}{2}(d_i^x-id_i^y)$ on all sites \footnote{We still assume $J_z$ cancels the antiferromagnetic Ising couplings arising from the second order perturbation to simplify the discussion.}, which is also gapped with nonzero Chern number, and the ground state now has non-abelian Ising topological order. 
More interestingly, the $\mathbb{Z}_2$  vortex has a zero mode which hosts a projective representation of $U(1)_z$ spin rotation symmetry with half quantum number $S^z_{\text{eff}}=\frac{1}{2}\sigma^z=\frac{\hbar}{2}$ in the low energy subspace $\mathcal{G}$.
Since the $f$-fermion has a nonzero Chern number $\nu=\pm1$ and its $U(1)$ electric charge is 
$S^z_{\text{eff}}=\hbar$, the $S^z_{\text{eff}}$ spin Hall conductance is quantized as $\sigma^s_{xy}=\nu\frac{\hbar}{2\pi}$ \cite{Yao2011}. As a result, if we insert a $\mathbb{Z}_2$ $\pi$-flux (or $-\pi$-flux equivalently) in a plaquette, an $S^z_{\text{eff}}=\pm\pi\times\sigma^s_{xy}=\pm\nu\frac{\hbar}{2}$ will accumulate around the $\mathbb{Z}_2$ vortex. 
This implies the existence of a zero mode with a fractionalized 
spin $S^z_{\text{eff}}=\frac{\hbar}{2}$, and the sign of the accumulated spin $S^z_{\text{eff}}=\pm\frac{\hbar}{2}$ corresponds to whether the zero mode is occupied or not. 
Remarkably, the 
existence of fractionalized $S^z_{\text{eff}}=\frac{\hbar}{2}$ in this model is direct evidence of spin fractionalization in the ground state \footnote{
It is an emergent fractionalization since the spin-$\frac{1}{2}$ quantum number is in the low energy subspace $\mathcal{G}$. In the 
lattice Hilbert space the deconfined spinons carry 
spin-1, 
but they 
can only be created by nonlocal string operators due to the incompatibility between the spin-$\frac{1}{2}$ quantum number and the local Hilbert space. 
Consequently the deconfined spinons have effective spin-1/2 quantum number in the low energy subspace $\mathcal{G}$. 
}.

Secondly, 
TRS can also be explicitly broken on the honeycomb lattice by adding a three-site term to $\hat{H}'_{S=1}$ with $J_z=2J_\text{eff}$: 
\begin{equation}
H_{h}=h \sum_{\substack{\langle i j\rangle\in\alpha \\ \langle j k\rangle\in\beta}} \boldsymbol{\epsilon}^{\alpha \beta \gamma}\left[\tau_{i}^{\alpha} \tau_{j}^{\gamma} \tau_{k}^{\beta}\right]\otimes (S^x_{i}  S^x_{k}+S^y_{i}  S^y_{k})^2,
\end{equation}
where $h\ll |J|$ and the three neighbouring sites $i,j,k$ are illustrated in Fig.~\eqref{fig:lat}. A nontrivial effective Hamiltonian in the ground state sector already exists in the first order perturbation of $h$:
\begin{eqnarray}
    &&\Delta H_{h,\text{eff}}=\hat{P}H_{h}\hat{P}\nonumber\\
&&~~~~~~=\frac{h}{2} \sum_{\langle i j\rangle, \langle j k\rangle} \hat{u}_{i j} \hat{u}_{j k}\left[2i(f^{\dagger}_if_k-f^{\dagger}_kf_i)-i d_{i}^{z} d_{k}^{z}\right],
\end{eqnarray}
which gaps all the Majorana fermions with nonzero Chern number and preserves the $U(1)_z$ spin rotation symmetry at the same time. 
The ground state 
has the Ising nonabelian topological order and fractionalized quantum number $S_{\text{eff}}^z=\frac{\hbar}{2}$ in the $\mathbb{Z}_2$ vortex. 
What's more, we anticipate the accumulated fractionalized charge $S_{\text{eff}}^z=\pm\frac{\hbar}{2}$ around the $\mathbb{Z}_2$ vortex will persist in the gapless phase as we continuously turn off the perturbation $h$ without changing the structure of zero modes.

{\it Large-S limit of the Yao-Lee model.}---To further 
explore new physics of the SU(2) symmetric spin-S Yao-Lee model, we 
now investigate the model's large-S limit 
which makes it analytically
tractable. To the leading order in the large-S expansion, we can treat the spin $\vec{S}_i$ as classical variables, and fractionalize the orbital DOF into Majorana fermions: $\vec{\tau}_i=i  \vec{c}_{i}c_{i}^0$. Now the large-S Yao-Lee model becomes a quadratic Hamiltonian: $\hat{H}_{S\to\infty} 
=\sum_{\langle i j\rangle } t_{ij}ic^0_ic^0_j$, where $t_{ij}=J \vec{S}_i\cdot \vec{S}_j\hat{u}_{ij}$ are effective hopping amplitudes, and $\hat{u}_{\ij}=ic^{\mu}_ic^{\mu}_j$ are static $\mathbb{Z}_2$ gauge fields. To find out the ground state, we need to identify the spin configuration $\{\vec S_i\}$ which can minimize the energy $E_0(\{\vec S_i\})$ of $\hat{H}_{\text{classical}}$.
With the help of Lieb's theorems \cite{PhysRevLett.107.066801,PhysRevLett.73.2158}, we find that 
the 
lowest ground state energy is 
achieved by co-linear spin ordering 
with infinite degeneracy: the states with $\vec{S}_i=\eta_i S\hat e$  ($\hat e$ being a common polarization axis but $\eta_i$ being $+1$ or $-1$ independently) 
are degenerate in energy 
as they can be connected through gauge transformations of $\hat{u}_{ij}$ \cite{SupMat}. 

In the next order of large-S expansion, we derived the  
effective 
model by incorporating quantum fluctuations of spins: $H_\textrm{eff}=g|J|S\sum_{\langle ij\rangle}(\eta_i\eta_j-1)$, from which it is clear that 
the degeneracy in large-S limit  
can be lifted by spin quantum fluctuations 
and a Neel order is favored in the ground state 
\cite{SupMat}. Coexisting with the spin Neel order, the orbital DOF fractionalizes into deconfined fermions with a Dirac cone. 
Due to the spontaneous breaking of SU(2) spin rotational symmetry, the spectrum here contains a  
Goldstone mode.  Results of the large-S limit imply a rich phase diagram by varying the spin quantum number. As there is no spin long-range order in the spin-$\frac{1}{2}$ Yao-Lee model, there should exist a critical spin $S_c$ separating the fractionalized spin $\vec{S}_i$ phases at small $S<S_c$ and ordered spin phases at large $S>S_c$, while the orbital $\vec{\tau}_i$ always fractionalize in these phases.

{\it Concluding remarks.}---Our work has revealed many physical implications of exact $\mathbb{Z}_2$ gauge structure on topological quantum liquids in non-integrable higher spin models, especially the previously largely unexplored integer spin ones. We have shown that the spin-S Yao-Lee model has QSOL ground state for any spin-S by constructing exact deconfined fermionic $\mathbb{Z}_2$ gauge charges without solving it. Further, in the minimal integer spin $S=1$ case, we analytically show the hidden deconfined nonabelian spinon with emergent fractionalized spin-$\frac{1}{2}$ quantum number in the $\mathbb{Z}_2$ vortex. The exact $\mathbb{Z}_2$ gauge structure enables us to provide a long sought after local spin Hamiltonian with the spin-1 bosonic Moore-Read Pfaffian topological order for future studies.

 {\it Acknowledgements:} H.Y. is grateful to Prof. Dung-Hai Lee for earlier collaborations on the Yao-Lee model. We also sincerely thank Linhao Li, Zijian Wang, Liujun Zou, and especially Zhou-Quan Wan for helpful discussions. This work is supported in part by the National Key R$\&$D Program of China under Grant No. 2021YFA1400100, NSFC under Grants No. 12347107 and No. 12334003 (Z. W., J. Y. Z., and H. Y.), the New Cornerstone Science Foundation through the Xplorer Prize (H. Y.), and by the Innovation Program for Quantum Science and Technology (grant No. 2021ZD0302502). Z.W. acknowledges the support in part from Shuimu fellowships at Tsinghua University.

\bibliographystyle{apsrev4-2}
\bibliography{references.bib}

\newpage
\widetext

\begin{center}
\textbf{\large Supplementary Material for 'Exact deconfined gauge structures in the higher-spin Yao-Lee model: \\a quantum spin-orbital liquid with spin fractionalization and non-Abelian anyons'}
\end{center}

\setcounter{equation}{0}
\setcounter{figure}{0}
\setcounter{table}{0}
\makeatletter
\renewcommand{\theequation}{S\arabic{equation}}
\renewcommand{\thefigure}{S\arabic{figure}}
\renewcommand{\bibnumfmt}[1]{[S#1]}
\renewcommand{\citenumfont}[1]{S#1}
\subsection{A. Derivation of the Hamiltonian in the parton representation}
In this section, we derive the parton representation $\hat{H}^f$ (Eq.(5) in the maintext) of the spin-S Yao-Lee model. The spin-S Yao-Lee model is:
\begin{equation}
\hat{H}=-\sum_{\mu=x,y,z} J_{\mu} \sum_{\langle i j\rangle \in \mu}[\vec{S}_i\cdot \vec{S}_j]\otimes[ \tau_{i}^{\mu} \tau_{j}^{\mu}],
\label{model}
\end{equation}
We first decompose $\vec{S}_i$  into $2S$ spin-$\frac{1}{2}$ operators $S_a^{\mu}$: $S_i^{\mu}=\sum_{a=1}^{2 S} S_{a,i}^{\mu}=\frac{1}{2}\sum_{a=1}^{2 S} \sigma_{a,i}^{\mu}$ with the local constraints: $
\sum_{\mu}\left(\sum_{a=1}^{2 S} S_{a,i}^{\mu}\right)^{2}=S(S+1)
$ . Then we introduce the Majorana representation of the orbital and spin as: 
\begin{equation}
\begin{aligned}
    &\tau^{\mu}_{i}=\frac{i}{2}\epsilon^{\mu\nu\rho}c_i^{\nu}c_i^{\rho},\quad \sigma^{\mu}_{1,i}=\frac{i}{2}\epsilon^{\mu\nu\rho}d^{\nu}_{1,i}d^{\rho}_{1,i},\\
    &\sigma^{\mu}_{a,i}= i  d^{\mu}_{a,i}d_{a,i}^0,\quad  a=2,3...2S.
    \end{aligned}
\end{equation}
The physical Hilbert space can be obtained under the following projections:
\begin{equation}
\begin{aligned}
&D_{i}=-i  d_{1,i}^{x} d_{1,i}^{y} d_{1,i}^{z}c_{i}^{x} c_{i}^{y} c_{i}^{z}=1, \\
&d_{a,i}^xd_{a,i}^yd_{a,i}^z d_{a,i}^0=1,\quad  a=2,3...2S
\end{aligned}
\end{equation}
The Hamiltonian Eq.\eqref{model} can be rewritten under this Majorana representation. We first rewrite the physical spin using the 2S Pauli matrices: $\hat{H}=-\frac{1}{4}\sum_{\mu,\langle i j\rangle \in \mu}  \sum_{a,b}J_{\mu}[\vec{\sigma}_{a,i}\cdot \vec{\sigma}_{b,j}]\otimes[ \tau_{i}^{\mu} \tau_{j}^{\mu}]$. Then the operator on each site can be written into fermions:
\begin{equation}
\begin{aligned}
  &\sigma^{\nu}_{a,i}\tau^{\mu}_i=id_{a,i}^0d^{\nu}_{a,i} d_{1,i}^{x}d_{1,i}^{y} d_{1,i}^{z}c_i^{\mu},\quad a=2,3...2S\\
    &\sigma^{\nu}_{1,i}\tau^{\mu}_i=id^{\nu}_{1,i}c_i^{\mu}
    \end{aligned}
\end{equation}
Now it is straightforward to rewrite the Hamiltonian in terms of the Majorana fermions:
\begin{equation}
\begin{aligned}
   H_f&=-\frac{1}{4}\sum_{\mu,\langle i j\rangle \in \mu}  J_{\mu}\hat{u}^{\mu}_{\langle ij\rangle}\sum_{\nu;a,b\neq1}i[(d^{\nu}_{1,i}+d^{\nu}_{a,i}d_{1,i}^{x}d_{1,i}^{y} d_{1,i}^{z}d_{a,i}^0)(d^{\nu}_{1,j}+d^{\nu}_{b,j}d_{1,j}^{x}d_{1,j}^{y} d_{1,j}^{z}d_{b,j}^0)]\\
   &=-\frac{1}{4}\sum_{\langle i j\rangle \in \mu}  J_{\mu}\hat{u}_{\langle ij\rangle}\sum_{a,b=2}^{2S}i{\vec D}_{a,i}\cdot{\vec D}_{b,j}, 
   \end{aligned}
\end{equation}
where  $\vec{D}_{a,i}=\vec{d}_{1,i}+\vec{d}_{a,i}d_{1,i}^{x}d_{1,i}^{y} d_{1,i}^{z}d_{a,i}^0$ is the composite field with $a=2,\cdots,2S$. The $\mathbb{Z}_2$ gauge fields $\hat{u}_{\langle ij\rangle}=-ic^{\mu}_{i}c_j^{\mu}$ are conserved quantities.

We want to further comment on the properties of the Hermitian composite fields $\vec{D}_{a,i}=\vec{D}_{a,i}^{\dagger}$. Although $\vec{D}_{a,i}=\vec{d}_i$, when spin $S=\frac{1}{2}$, are fermionic fields, they are no longer standard fermionic fields when spin $S>\frac{1}{2}$. Two composite operators from different lattice sites indeed anti-commute with each other: $\{D^{\mu}_{a,i},D^{\nu}_{b,j}\}=0$ when $i\neq j$ for any $a,b,\mu,\nu$, but two same site operators $D^{\mu}_{a,i}$ and $D^{\nu}_{b,i}$ ($D^{\mu}_{a,i}\neq D^{\nu}_{b,i}$) neither commute nor anti-commute with each other.
\subsection{B. Proof of the ground state subspace}
In this section, we prove the low energy ground state sector $\mathcal{G}$ of $\hat{H}_0$ in the maintext is given by all the ground state of $\hat{H}_0$ with nonzero $S^z$. We complete the proof through proof by contradiction.

We suppose $|\psi(\{S^z\})\rangle$ is a ground state of $\hat{H}_0$ with at least one $S_i^z=0$, and we prove that its energy must be strictly higher than the real ground state. Actually we only need to consider the states with a single zero $S_i^z$. Indeed, if there exist at least two zero $S^z$: $S_i^z=S_j^z=0$ and $j\neq i$ in a state $|\psi^{\prime}\rangle$, then we can add a term $\Delta H_j=\eta_j\sum_{k\in \langle jk\rangle}S_k^z\otimes \tau_j^{\mu}\tau_k^{\mu}$ to $\hat{H}_0(\{S^z\}^{\prime})$, where $\eta_j=\pm 1$ is a $\mathbb{Z}_2$ variable to make the expectation value $\langle\psi^{\prime}|\Delta H_j|\psi^{\prime}\rangle\leq0$ and $\{S^z\}^{\prime}$ is the $S^z$ configuration of $|\psi^{\prime}\rangle$ . Now the ground state energy of the Hamiltonian $\hat{H}_0(\{S^z\}^{\prime})+\Delta H_j$ cannot be higher than the energy of $|\psi^{\prime}\rangle$, but the new ground state has one less zero $S^z$. We can continue this process until there is only one zero $S_i^z$ left, and we only need to prove the energy of the final state $|\psi(\{S^z\})\rangle$ is strictly higher than the real ground state of $\hat{H}_0$. Meanwhile, an immediate corollary is that the ground states of any $\hat{H}_0(\{S^z\neq 0\})$ must belong to the real ground states sector of $\hat{H}_0$, since we can just continue the addition of $\Delta H$ until no zero $S^z$ is left. 

Now if $\langle\psi(\{S^z\})|\sum_{j\in \langle ij\rangle}S_j^z\otimes \tau_i^{\mu}\tau_j^{\mu}|\psi(\{S^z\})\rangle\neq 0$, then we can find a nonzero $\eta_i$ and the ground state of $\hat{H}(\{S^z\})+\Delta H_i$ has strictly lower energy and then the proof is complete. So the only corner case left is $\langle\psi(\{S^z\})|\sum_{j\in \langle ij\rangle}S_j^z\otimes \tau_i^{\mu}\tau_j^{\mu}|\psi(\{S^z\})\rangle= 0$, and we can prove that this contradicts the assumption that $|\psi(\{S^z\})\rangle$ is the ground state of $\hat{H}_0$. We write $|\psi(\{S^z\})\rangle$ in the Majorana representation: $|\psi(\{S^z\})\rangle=\Pi_i \hat{D}_i |\{u_{\langle ij\rangle}\}\rangle \otimes |c^0\rangle$, where $ \hat{D}_i=\frac{1+c_i^xc_i^yc_i^zc_i^0}{2}$ is the projection on site $i$. If $\langle\psi(\{S^z\})|\sum_{j\in \langle ij\rangle}S_j^z\otimes \tau_i^{\mu}\tau_j^{\mu}|\psi(\{S^z\})\rangle= 0$, then $|\psi(\{S^z\})\rangle$ is also the ground state of $\hat{H}_0^{\prime}=\hat{H}_0(\{S^z\})+\sum_{j\in \langle ij\rangle}S_j^z\otimes \tau_i^{\mu}\tau_j^{\mu}$. The $S^z$ configuration of $\hat{H}_0^{\prime}$ has no zero since $S_i^z=1$, so in the Majorana representation of $\hat{H}_0^{\prime}$, we can just take the $\mathbb{Z}_2$ gauge field configuration to make the remaining itinerant fermion Hamiltonian preserve all the lattice symmetries, which has only one ground state $|c^0\rangle$ with one Dirac cone. As a result, $\langle\psi(\{S^z\})|\sum_{j\in \langle ij\rangle}S_j^z\otimes \tau_i^{\mu}\tau_j^{\mu}|\psi(\{S^z\})\rangle$ is just the expectation value of the summation of three nearest neighbour hoppings: $\sum_{j\in \langle ij\rangle}S_j^zu_{\langle ij\rangle}\langle c^0|ic^0_ic^0_j|c^0\rangle$, which must be nonzero since $\langle c^0|ic^0_ic^0_j|c^0\rangle$ are the same on all bonds, and $S_j^zu_{\langle ij\rangle}$ are three $\mathbb{Z}_2$ variables whose summation cannot be zero.
\subsection{C. Derivation of the effective Hamiltonian through degenerate perturbation}
In this section, we derive the effective Hamiltonian of Eq. (6) in the main text through a degenerate perturbation to the second order of a. Since we have identified the zeroth order ground state sector $\mathcal{G}$ in the previous section, which are the ground states of the unperturbed Hamiltonian $\hat{H}_0=-J\sum_{\langle i j\rangle \in \mu}S_i^zS_j^z\otimes \tau_{i}^{\mu} \tau_{j}^{\mu}$ with all the $S^z$ nonzero. So the spin degrees of freedom on each site is an effective spin-$\frac{1}{2}$, and we denote this spin-$\frac{1}{2}$ as: $\hat{P}S^z_i\hat{P}=\sigma^z_i$, where $\hat{P}$ is the projection to the low-energy ground state sector $\mathcal{G}$. A nontrivial effective Hamiltonian exists at the second order perturbation:
\begin{equation}
    \hat{H}_{\text {eff }}=\hat{P}\hat{H}^0\hat{P}+\hat{P} V\left(\frac{1}{E_{0}-\hat{H}_{0}} \hat{Q} V\right) \hat{P},
\end{equation}
where $E_0$ are the ground state energy of $\hat{H}_0$; $\hat{Q}=1-\hat{P}$ is the projection to the excited states, and $V=-aJ\sum_{\langle ij\rangle \in \mu}(S_i^xS_j^x+S_i^yS_j^y)\otimes[ \tau_{i}^{\mu} \tau_{j}^{\mu}]$ is the perturbation with $0<a\ll1$. We assume the energy gap between $\mathcal{G}$ and all the low-lying intermediate states are approximately the same $\Delta |J|$, where $\Delta$ is an $O(1)$ dimensionless postive number, then we can write down a simple form of $\hat{H}_{\text {eff }}$:
\begin{equation}
\begin{aligned}
    \hat{P} V\left(\frac{1}{E_{0}-\hat{H}_{0}} Q V\right) \hat{P}&=-\frac{a^2 |J|}{\Delta} \hat{P}\left[\sum_{\langle ij\rangle} (S_i^xS_j^x+S_i^yS_j^y)^2\right]\hat{P}\\
    &=-\frac{a^2|J|}{\Delta} \sum_{\langle ij\rangle}\hat{P}\left[ (S_i^xS_j^x)^2+(S_i^yS_j^y)^2+S_i^xS_i^y\otimes S_j^xS_j^y+S_i^yS_i^x\otimes S_j^yS_j^x\right]\hat{P}\\
    &=- \frac{a^2 |J|}{4\Delta }\sum_{\langle ij\rangle}\left[(\sigma_i^x+1)(\sigma_j^x+1)+(\sigma_i^x-1)(\sigma_j^x-1)+(\sigma_i^y+i\sigma_i^z)(\sigma_j^y+i\sigma_j^z)+(\sigma_i^y-i\sigma_i^z)(\sigma_j^y-i\sigma_j^z)\right]\\
    &=- \frac{a^2|J|}{2\Delta }\sum_{\langle ij\rangle}\sigma_i^x\sigma_j^x+\sigma_i^y\sigma_j^y-\sigma_i^z\sigma_j^z+\text{const.},
    \end{aligned}
\end{equation}
where is exactly the Eq. (8) in the main text with $J_{\text{eff}}=\frac{a^2|J|}{2\Delta }$.

\subsection{D. Large-S limit of the spin-S Yao-Lee model}
The spin-S Yao-Lee model $\hat{H}=J\sum_{\langle i j\rangle \in \mu} [\vec{S}_i\cdot \vec{S}_j]\otimes[ \tau_{i}^{\mu} \tau_{j}^{\mu}]$ is analytically solvable in a controlled way in limit of large-S ($S\rightarrow\infty$). 
In the leading order of large-S expansion, the spin degrees of freedom (DOF) $\vec{S}_i$ are just classical variables subject to the constraints: $| \vec{S}_i|=S$. To find out the ground state subspace, we fractionalize the orbital DOF into four Majorana fermions: $\vec{\tau}_i=i\vec{c}_ic^0_i$, and to the leading order of large-S expansion the Yao-Lee model becomes a quadratic fermion Hamiltonian:
\begin{equation}
    \hat{H}_{\text{classical}}=\sum_{\langle i j\rangle } t_{ij}ic^0_ic^0_j,
\end{equation}
where the nearest neighbour hopping amplitudes $t_{ij}=J \vec{S}_i\cdot \vec{S}_j\hat{u}_{ij}$, and $\hat{u}_{\ij}=ic^{\mu}_ic^{\mu}_j$ are static $\mathbb{Z}_2$ gauge fields on the bond $\langle ij\rangle$. 

\begin{figure}[b]
  \includegraphics[width=5cm]{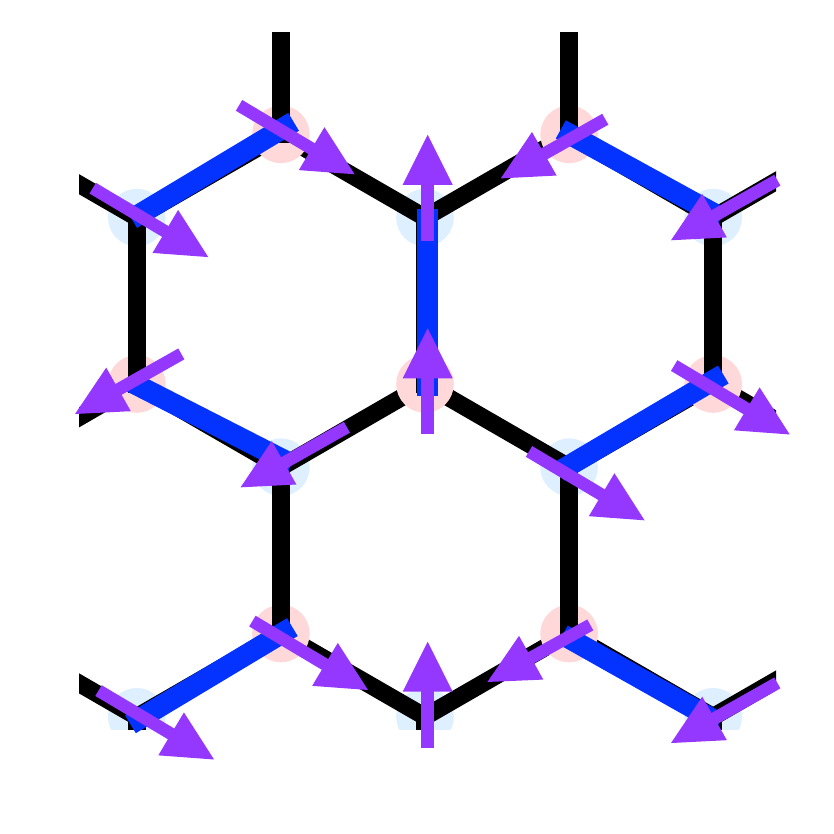}
 \caption{A possible $\sqrt{3}\times\sqrt{3}$ spin spiral order in the large-S limit of the Yao-Lee model. The purple arrows represent the spin order, and the blue and black bonds are the strong and weak bonds of Majorana hoppings, respectively.}
 \label{fig:larg}
\end{figure}

Here we focus on the honeycomb lattice. The Hamiltonian $\hat{H}_{\text{classical}} $ is reminiscent of Lieb's theorem on the phonon-coupled Hubbard model \cite{PhysRevLett.107.066801} and the generalized Hofstadter-Hubbard model \cite{PhysRevLett.73.2158}. According to Lieb's theorem, the ground state sector of $ \hat{H}_{\text{classical}}$ should lie in the zero flux sector, and the most general hopping configuration $\{t_{ij}\}$ is uniform or spontaneously breaks translational symmetry with the Kekul\'{e} pattern. A uniform $\{t_{ij}=t\}$ configuration can also be viewed as a special case of Kekul\'{e} pattern with zero Kekul\'{e} distortion. Formulated in the physical spin-orbital DOF, all the co-linear spin order configurations have the same energy, and the $c^0$ fermions have a Dirac cone. Nevertheless, a $\sqrt{3}\times \sqrt{3}$ co-planar $120^{\circ}$ spin spiral order can gap out the Dirac cone, since the resulting hopping $t_{ij}$ has a nonzero Kekul\'{e} distortion. The $\sqrt{3}\times \sqrt{3}$ spin spiral order is illustrated in Fig. \eqref{fig:larg}.

As a result, we consider a set of classical spin configuration as follows: $S^z_i=S\sin{\theta}$ represents the uniform component, while $S^{+}_{i,A}=S\cos{\theta}e^{-i \mathbf{K}\cdot \mathbf{r_i}},S_{i,B}^{+}=S\cos{\theta}e^{ i \mathbf{K}\cdot \mathbf{r_i}}$ contains the  Kekul\'{e} distortion, where $\mathbf{K}=(\frac{4\pi}{3},0)$ and the lattice distance between the nearest AA (or BB) sites is set to be 1. The parameter $\theta$ can tune the relative weights between these two patterns, and the optimal $\theta$ can be arrived by minimizing the ground state energy. We would like to remark that we always consider the zero flux sector of the hopping $t_{ij}$ in the energy optimization by tuning the $\mathbb{Z}_2$ gauge fields $u_{\langle ij\rangle}$, which is guaranteed by Lieb's theorem to obtain the lowest energy state.

In the fermionic parton representation, the hopping amplitude of the strong bond is $t_s=JS^2(\sin^2{\theta}+\cos^2{\theta})=JS^2$, and the weak bond gives: $t_{w}=|-\frac{J}{2}S^2\cos^2{\theta}+JS^2\sin^2{\theta}|=JS^2|\frac{1}{4}-\frac{3}{4}\cos{2\theta}|$. According to the Hellmann-Feynman theorem, we obtain the following:
\begin{equation}
   0=\frac{\partial E_g(\theta)}{\partial_\theta}=\langle\frac{\partial \hat{H}(\theta)}{\partial_\theta}\rangle=\pm\frac{3\sin{2\theta}}{2}\sum_{\langle ij\rangle\in\text{weak bond}}\langle ic^0_ic^0_j\rangle. 
   \label{classical}
\end{equation}

Three states obviously satisfy the above equation:

(i) co-linear spin order, which corresponds to $\cos\theta=0$; 

(ii) $120^{\circ}$ spin spiral order, which corresponds to $\sin\theta=0$;

(iii) Zero hopping on the weak bonds, which corresponds to $\cos(2\theta)=\frac{1}{3}$. 

\begin{figure}[t]
  \includegraphics[width=13cm]{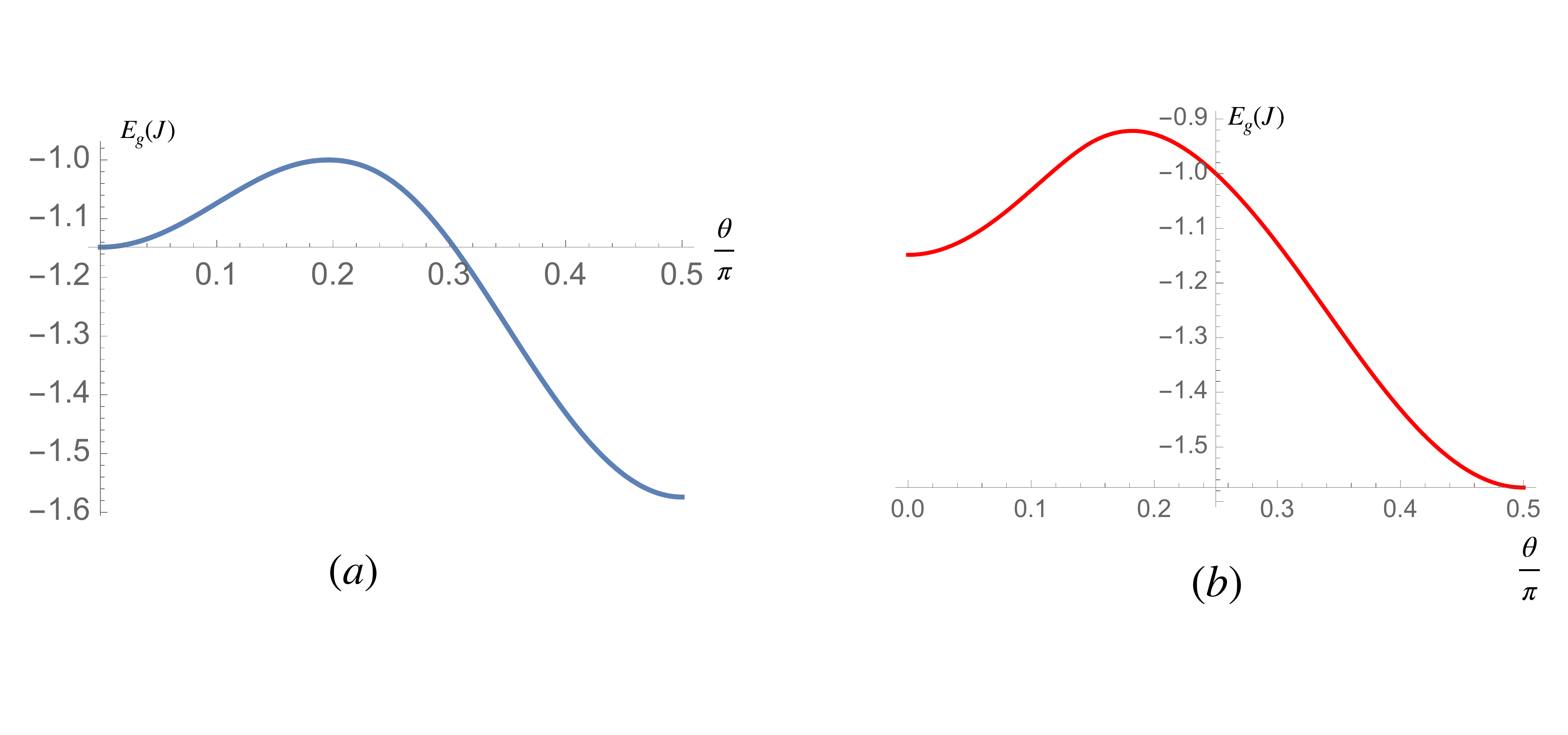}
 \caption{ The ground state energy dependence on the variational parameter $\theta$ in two classes of trial wavefunctions. $\theta=\frac{\pi}{2}$ corresponds to co-linear spin order with all $\vec{S}_i$ in the $z$-direction, and $\theta=0$ corresponds to a purely $\sqrt{3}\times\sqrt{3}$ spin order in the xy plane, as is illustrated in Fig. \eqref{fig:larg}. We always consider the zero-flux sector of the $c^0$ partons in Eq. \eqref{classical}, as required by the Lieb's theorem. (a) Ground state energy with ferromagnetic $S_i^z$ components. (b) Ground state energy with anti-ferromagnetic $S_i^z$ components. }
 \label{fig:energy}
\end{figure}

The numerical results of the energy as a function of $\theta$ are illustrated in Fig. \eqref{fig:energy}(a), from which we can see that only these three states satisfy the stationary condition $\frac{\partial E_g(\theta)}{\partial \theta}=0$, and co-linear spin order gives rise to the lowest ground state energy. With the presence of co-linear spin order, the orbital degrees of freedom fractionalize into a Dirac cone. In Fig. \eqref{fig:energy}(b), we consider another set of variational wave functions, which have antiferromagnetic $S_i^z$ components and the $S_i^x,S_i^y$ components still take the $\sqrt{3}\times\sqrt{3}$ pattern. As we can see from the results in Fig. \eqref{fig:energy} (b), the lowest energy state is still reached by co-linear spin order.

We note that the ground state has infinite degeneracy in the leading order of large-S expansion, since the spin $S_i$ on each site can be independently spin up or down along the ordered direction, which means there is a local $\mathbb{Z}_2$ variable $\eta_i=\pm 1$ on each site in the ground state subspace. We further show that the inclusion of quantum fluctuation of $S_i$ picks the Neel order as the ground state to the next order in the $\frac{1}{S}$ expansion, which is similar to the 'order by disorder' mechanism.  
We include the quantum fluctuation of spin using a real space second-order perturbation as in Refs.  \cite{PhysRevLett.61.629,PhysRevB.48.7227,PhysRevLett.113.237202,MWLong1989,PhysRevLett.118.147204,Rousochatzakis2018}. 
We denote the spin operator on each site as: $\vec{S}_i=\hat{S}_i^x \hat{e}_i^x+\hat{S}_i^z \hat{e}_i^y+\hat{S}_i^z \hat{e}_i^z$, where $\hat{e}_i^\mu,\mu=x,y,z$ are local spin basis; $\hat{e}_i^z$ is the ordering direction of spin in the classical limit with $\hat{e}^z_i\cdot\hat{e}^z_j=\eta_i\eta_j$, where $\eta_i,\eta_j$ are local $\mathbb{Z}_2$ variables, and $\hat{e}^{x}_i \cdot\hat{e}^{x}_j=\eta_i\eta_j$ and $\hat{e}^{y}_i \cdot\hat{e}^{y}_j=1$. We further divide the Hamiltonian $\hat H$ of the Yao-Lee model into $\hat{H}_0+V$, where the unperturbed Hamiltonian $\hat{H}_0= J\sum_{\langle ij\rangle\in\mu}\hat{S}_i^z\hat{S}_j^z\eta_i\eta_j \hat{\tau}^{\mu}_i\hat{\tau}_j^{\mu}$. Using the standard Holstein-Primakoff transformation in the large-S limit: $\hat{S}_i^z=S-\hat{a}^{\dagger}_i\hat{a}_i=S-\hat n_i, \hat{S}_i^{+}\approx\sqrt{2S}\hat{a}_i,\hat{S}_i^-\approx\sqrt{2S}\hat{a}_i^{\dagger}$, and under the linear magnon approximation, the unperturbated $\hat{H}^0$ becomes:
\begin{equation}
\begin{aligned}
     \hat{H}_0&= J\sum_{\langle ij\rangle\in\mu}(S-\hat{n}_i)(S-\hat{n}_j)\eta_i\eta_j \hat{\tau}^{\mu}_i\hat{\tau}_j^{\mu}\\
     &\approx\sum_{\langle ij\rangle\in\mu} \left[JS^2-JS(\hat{n}_i+\hat{n}_j)\right]\eta_i\eta_j\hat{\tau}_i^{\mu}\hat{\tau}_j^{\mu}.
\end{aligned}
\end{equation}
The perturbation part is: $V=J\sum_{\langle ij\rangle\in\mu} (\hat{S}_i^x\hat{S}_j^x\eta_i\eta_j+\hat{S}_i^y\hat{S}_j^y)\hat{\tau}_i^{\mu}\hat{\tau}_j^{\mu}$. At the second order perturbation (the $\frac{1}{S}$ expansion), we have the effective Hamiltonian of the $\eta_i$ variables as:
\begin{equation}
    \begin{aligned}
        H_{\text{eff}}&= -J^2\hat{P}\sum_{\langle ij\rangle}(\hat{S}_i^+\hat{S}_j^+\eta_i\eta_j-\hat{S}_i^+\hat{S}_j^+)\frac{1}{\tilde g |J|S}(\hat{S}_i^-\hat{S}_j^-\eta_i\eta_j-\hat{S}_i^-\hat{S}_j^-)\hat{P}\\
        &=g|J|S\sum_{\langle ij\rangle}(\eta_i\eta_j-1),
    \end{aligned}
\end{equation}
where $\hat{P}$ is the projection to the co-linear spin order subspace and $g=8/\tilde g$ is a positive constant of order one (here as the energy gap of all relevant intermediate states is proportional to $S|J|$, it can be approximated by $\tilde g S|J|$, where $\tilde g$ is a positive constant of order one). Note that only the configurations $\eta_i\eta_j=-1$ can gain energy from the second order perturbations, while the $\eta_i\eta_j=1$ configurations actually do not have this second order perturbation process. 
As a result, we can immediately arrived at the conclusion that the spin quantum fluctuation tends to favor a spin Neel order in the ground state irrespective of the sign of the Yao-Lee coupling $J$. Now, in the presence of spin Neel order the orbital degrees of freedom fractionalize into a Dirac cone, which can be seen from a parton representation $\vec{\tau}_i=i\vec{c}_ic^0_i$.


\end{document}